\documentclass[amsmath,aip,graphicx,reprint]{revtex4-1}

\usepackage{graphicx,amssymb,siunitx}
\DeclareSIUnit\gauss{G}

\draft

\begin{document}

\title{Finite-amplitude RF heating rates for magnetized electrons in neutral plasma} 

\author{John M Guthrie}

\author{Jacob L Roberts}
\email[]{jacob.roberts@colostate.edu}

\affiliation{Department of Physics, Colorado State University, Fort Collins, Colorado 80523, USA}

\date{\today}

\begin{abstract}
A theoretical model is developed and evaluated using a Vlasov-Poisson treatment to calculate radiofrequency (RF) electric field heating rates for magnetized electrons in neutral plasma when the magnetic and electric field directions are colinear and when the RF is of sufficiently high frequency. This calculation reduces to the theory for magnetized longitudinal AC conductivity introduced by Oberman and Shure [C.~Oberman and F.~Shure, Phys. Fluids {\textbf{6}}, 834-838 (1963)] in the low-amplitude limit when the electron oscillation velocity is much less than the thermal velocity. For electron coupling strengths $\Gamma=0.15$--$0.015$ and RF fields accessible to ultracold neutral plasma experiments, the model predicts mild variations in heating rate of order unity across magnetization strengths spanning orders of magnitude. The predicted effect of including a BGK-type collisional relaxation term in the Vlasov equation reduces the heating rate by 5\% or less across magnetizations.
\end{abstract}


\maketitle

Magnetic fields are common in laboratory and astrophysical plasmas.  As a magnetic field becomes stronger, plasma transport, then screening, and then collision properties are predicted to be substantially affected.\cite{Baalrud2017}  Collisions that are affected at sufficiently strong magnetic fields include electron-ion collisions, and these collisions are a fundamental property of plasmas that determine quantities such as diffusion and other transport rates,\cite{Baalrud2012,Dubin2014,Stanton2016,Baalrud2017,Shaffer2019} thermalization rates,\cite{Ichimaru1970,Steck1994,Beutelspacher2004,Chen2016,Jiang2020} and stopping power.\cite{Honda1963,MayCramer1970,Peter1991,NersisyanBook,Hu2009,Hu2010,Nersisyan2011,Nersisyan2011Fluc,Nersisyan2011Laser,Baalrud2020} The electron component of a two-component (electron and ion) plasma is generally much easier to magnetize than the ion component since the electron mass is much smaller.  While one-component electron plasmas have been used to investigate temperature equilibration rates as a function of electron magnetization,\cite{Glinsky1992,Beck1996} comparable systematic studies of electron-ion collision rates as a function of electron magnetization have not been performed (which is not to discount electron cooling experiments that are sensitive to electron-ion collision rates\cite{Steck1994,Beutelspacher2004}---however it is hard to systematically vary the degree of electron magnetization in these systems, and they also have large temperature anisotropies that limit the general applicability of their results). The purpose of this work is to theoretically examine the predicted effects of electron magnetization on electron-ion collision rates in a plasma system where they can be measured over a wide range of electron magnetization, namely in ultracold neutral plasmas (UNPs).\cite{Killian1999,Killian2007} This is done to produce predictions that can be tested, to guide the selection of parameters where the most useful experiments can be performed, and to evaluate the likely importance of greater-than-first-order effects for achievable experimental conditions. 

In particular, we focus on conditions consistent with low-density UNPs where high degrees of magnetization are most easily obtained.\cite{Killian2007} This in part determined the range of electron strong coupling parameter that was investigated through considering values most commonly present in these UNPs, while avoiding more strongly coupled conditions that are beyond the scope of this work.

One way that electron-ion collision rates can be measured is through the electron heating rate in plasmas subjected to an oscillating electric field. For laser-driven plasmas, such heating occurs through inverse bremsstrahlung absorption.\cite{Langdon1980,Decker1994,Liu1994,Pfalzner1998,Kundu2015,Sedaghat2015,Farrashbandi2020} Because of the much lower density in UNPs, the heating can be applied through radiofrequency (RF) rather than optical fields. This heating occurs through the finite AC conductivity of these plasmas, which is itself related to the electron-ion collision rate. The use of RF fields simplifies the system since the field absorption is not significant,\cite{Rolston2012} and so the absorption of the applied field does not have to be modeled, unlike in the optical case. The focus of the work in this article is on making predictions for such finite AC conductivity heating for plasmas with UNP parameters. The UNP parameters investigated border on the strongly coupled region for the electrons, with $\Gamma$ up to 0.15 investigated in this work (and larger values predicted to be possible in UNPs).\cite{Chen2017} In addition, degrees of electron magnetization into the ``extreme'' region are studied since it is possible to achieve these extreme degrees of magnetization in UNPs with only mild laboratory fields.

Predictions for such AC conductivity heating in plasmas with magnetized electrons exist in the literature.\cite{ObermanShure1963,Chen1974,Matsuda1981} However, these predictions occur in the context of assumptions that lead to practical difficulties for UNP experiments. The electric fields in Refs.~\onlinecite{ObermanShure1963,Chen1974,Matsuda1981} are assumed to be smaller than a limiting field that itself is too small for many possible UNP experiments given realistic signal-to-noise considerations. Predictions for nonlinear effects on the AC conductivity due to large oscillation amplitudes exist for the unmagnetized case.\cite{Silin1965,Decker1994,Pfalzner1998} While there are theories that investigate nonlinear electric field absorption as a function of magnetic field strength, they do not examine how the magnetic field affects the nature of the nonlinearity over a range of magnetizations.\cite{Sedaghat2015} It is thus necessary to evaluate whether or not substantial changes in the heating rate across electron magnetization strengths are expected for electric field amplitudes above the limits identified in Refs.~\onlinecite{ObermanShure1963,Chen1974,Matsuda1981}. 

In light of these considerations, we have derived predictions from a theory based on linearized Vlasov and Poisson equations\cite{NersisyanBook} where the main modification is to alter the theory in a straightforward way to remove these same electric field limitations. Thus, these predictions should be useful for evaluating how finite amplitude electric fields could impact UNP RF heating measurements. The theory that is presented in this work is comparable to those developed for magnetized stopping power calculations\cite{Honda1963,MayCramer1970,Peter1991,Deutsch2005,NersisyanBook,Hu2009,Hu2010,Nersisyan2011,Nersisyan2011Fluc} and unmagnetized AC conductivity predictions\cite{Silin1965,Decker1994,Kundu2015} and combines the ability to treat applied electric fields greater than low-field limits in plasmas with a high degree of magnetization. 

Since this article is concerned with UNPs, a brief review is presented here. UNPs are created through photoionizing either ultracold atoms,\cite{Killian1999,Gallagher2000,Raithel2008,Wilson2013,Killian2019,Bergeson2019} or atoms or molecules in a supersonic expansion.\cite{Grant2008,Grant2014} After photoioniztion, some electrons escape from the plasma region until a space charge develops due to the more massive ions, confining the remaining electrons to form a quasi-neutral electron-ion plasma with very low species temperatures, low density, and a finite spatial extent. In most UNPs, the ions are unconfined and so will eventually expand after formation. It is possible to conduct experiments shortly after the UNP formation where this expansion is not significant.\cite{Chen2017}

In this work, we concern ourselves with the AC conductivity heating for oscillations in the longitudinal direction (i.e. the oscillating electric field and applied magnetic field directions are parallel) as that is the easiest geometry available in experimental UNP systems.\cite{Zhang2008} In this article we review the Vlasov-Poisson equation-based theory and its main features, compare it to the predictions of Ref.~\onlinecite{Matsuda1981} in the appropriate limits, and show predicted AC conductivity heating rates as a function of magnetic field and electric field amplitude as well as the applied RF frequency. We also comment on whether or not adding a collision term\cite{Nersisyan2011} to the Vlasov-Poisson equations is expected to make a significant difference to the predicted heating rate. We find agreement between this theory and the predictions of Ref.~\onlinecite{Matsuda1981} in the appropriate limits and that the addition of collisions is significant at only the few percent level for the conditions considered. The work presented here should thus be useful for predictions of RF heating relevant for UNP conditions at the several to tens of percent level of precision without needing to include corrections from the absence of a collisional term in the fundamental equations.  

\section{Vlasov-Poisson-based linear response RF heating rate}

The underlying equations for our calculations of the heating rate are the linearized Vlasov-Poisson equations,\cite{NersisyanBook}
\begin{eqnarray}
\frac{\partial f_{\nu1}}{\partial t} + \mathbf{v}\cdot\frac{\partial f_{\nu1}}{\partial \mathbf{r}} &+& \frac{q_\nu}{m_\nu}\left(\mathbf{v}\times\mathbf{B}\right)\cdot \frac{\partial f_{\nu1}}{\partial \mathbf{v}} \nonumber\\
 &-&\frac{q_\nu}{m_\nu}\frac{\partial \Phi}{\partial \mathbf{r}}\cdot\frac{\partial f_{\nu0}}{\partial \mathbf{v}} = \left(\frac{\partial f_\nu}{\partial t}\right)_{coll} \label{eq:linVP} \\
\epsilon_0\nabla^2\Phi = -\rho_\mathrm{i}(\mathbf{r},&t)& - \sum_\nu n_\nu q_v \int\mathrm{d}\mathbf{v}f_{1\nu}(\mathbf{r},\mathbf{v},t),
\end{eqnarray}
where $\rho_\mathrm{i}(\mathbf{r},t) = q\delta(\mathbf{r}-\mathbf{r}_\mathrm{i}(t))$ represents the charge density of a heavy particle, $f_\nu = f_{\nu0} + f_{\nu1}$ is the perturbed phase-space distribution for plasma species $\nu$ with uncorrelated equilibrium distribution $f_{\nu0}$, and $n_\nu = \int f_\nu \mathrm{d}\mathbf{v}$. For most of our predictions, we use the collisionless Vlasov equation by ignoring the collision term on the right-hand side of Eq.~(\ref{eq:linVP}). Later in this work collisions will be included approximately through a relaxation term\cite{Nersisyan2011} to provide an estimation of the possible size of the effect from a collision term. 

From these equations, the techniques used in magnetized stopping power theory\cite{Honda1963,MayCramer1970} presented in Ref.~\onlinecite{NersisyanBook} and unmagnetized AC heating rates in Ref.~\onlinecite{Decker1994,Kundu2015} are adapted to produce magnetized RF heating rate predictions. This method has been applied for stopping power in magnetized-electron two-component plasmas in the context of a heavy projectile of constant velocity,\cite{NersisyanBook,Hu2009} plasmas irradiated by laser field,\cite{Nersisyan2011Laser} and plasmas subject to electric field fluctuations.\cite{Nersisyan2011Fluc}

Here we consider the case of a heavy ion oscillating sinusoidally in the plasma at a selected amplitude parallel to the magnetic field. In this work, the plasma ions are assumed to be infinitely massive and uncorrelated.\cite{Matsuda1981} The potential $\Phi$ is equal to
\begin{eqnarray}
\nonumber\Phi(\mathbf{r},t)\,&=& \frac{q}{(2\pi)^4\epsilon_0}\int\mathrm{d}\mathbf{k}\int_{-\infty}^{\infty}\mathrm{d}\omega\frac{\exp\left(i\mathbf{k}\cdot\mathbf{r}-i\omega t\right)}{k^2\epsilon(\mathbf{k},\omega)}\\
&\times& \int_{-\infty}^{\infty}\mathrm{d}t'\exp\left[i\omega t'-i\mathbf{k}\cdot\mathbf{r}_\mathrm{i}(t')\right]
\label{eq:PhiEqn}
\end{eqnarray}
where $\epsilon(\mathbf{k},\omega)$ is the dielectric response function for electron-ion plasma in a uniform magnetic field.\cite{NersisyanBook} From $\Phi$ the resulting force on the ion during its oscillation can be calculated, and given the oscillation the resulting work on the ion can be determined. This procedure is related to the RF heating rate of electrons oscillating relative to ions through a reference frame transformation.\cite{Decker1994}

The RF field in these calculations has amplitude $E_0$ and frequency $\omega_{RF}$ so that $\mathbf{E}(t) = E_0\hat{\mathbf{z}}\sin(\omega_{RF}t + \phi)$, where $\phi$ is an arbitrary phase we may set to zero. This induces an oscillation of the electrons of amplitude $\mathbf{r}_\mathrm{i}(t) = a_0\hat{\mathbf{z}}\sin\omega_{RF}t$, $a_0 = qE_0/m\omega_{RF}^2$
in the high-frequency ($\omega_{RF} \gg \omega_p$) limit.
At frequencies closer to the electron plasma frequency, $\omega_p$, the electron center-of-mass response to an oscillating field can be taken into account in determining the amplitude, with the density and geometry of the plasma playing a role in the resulting oscillation amplitude.\cite{Wilson2013}

The work done on the plasma by the oscillating ion can be determined by substituting $\mathbf{r}_\mathrm{i}(t)$ into (\ref{eq:PhiEqn}) and calculating $S(t) = \mathbf{v}_\mathrm{i}\cdot e\vec{\nabla}\Phi/v_\mathrm{i}$, where $v_i = |\mathbf{v}_\mathrm{i}| = |\dot{\mathbf{r}}_\mathrm{i}|$ is the velocity amplitude associated with the oscillatory motion. For a constant $v_\mathrm{i}$, $S(t)$ is the stopping power of a plasma for an ion. Integrating $S(t)$ over a single RF oscillation cycle produces the work done on electrons
\begin{eqnarray}
\nonumber\Delta\mathcal{E}\, &=& \frac{2e^2}{\pi\epsilon_0}\int_0^{k_{max}}\!\!\mathrm{d}k\int_0^{\pi/2}\!\!\mathrm{d}\theta\sin\theta\sum_{n=1}^{\infty}nJ_n(ka\cos\theta)^2\\
&\times&\left[\mathrm{Im}\frac{-1}{\epsilon(\mathbf{k},n\omega_{RF})}\right].
\end{eqnarray}
We have used a number of Bessel function relationships in the derivation of the above expression, such as the Jacobi-Anger identity $\exp\left(-iz\sin\theta\right) = \sum_{n=-\infty}^{\infty} J_n(z)\exp(-in\theta)$.\cite{DLMF} The cutoff parameter $k_{max}$ in the $\mathrm{d}k$ integral is included to avoid logarithmic divergences. It embodies the failure of the linear response theory in the small impact parameter, hard-angle scattering limit. We chose $k_{max} = r_{min}^{-1} = 4\pi\epsilon_0k_\mathrm{B}T_\mathrm{e}/e^2$ for this cutoff, where $r_{min}$ is a characteristic collision length scale related to the classical distance of closest approach.

The magnetic field-dependent energy change $\Delta\mathcal{E}$ can be related to a scaled, dimensionless heating rate $H\equiv\eta/\omega_p$ assuming an idealized form $\mathrm{d}\mathcal{E}/\mathrm{d}t = \eta m_\mathrm{e} v_\mathrm{i}^2$ where $\mathcal{E}$ is the electron kinetic energy and $\eta$ is a rate and proportionality factor set by $\Delta \mathcal{E} = \pi \eta m_\mathrm{e} a^2 \omega_{RF}$. We can express $H$ in terms of the coupling strength $\Gamma$ and scaled parameters
\begin{eqnarray*}
\begin{tabular}{r@{$\,\leftrightarrow\,$}l@{$\quad$}r@{$\,\leftrightarrow\,$}l}
$\mathbf{k}'$ & $\lambda_\mathrm{D}\mathbf{k}$ & $\omega'$ & $\omega_{RF}/\omega_p$\\
$a'$ & $a/\lambda_\mathrm{D}$ & $\beta$ & $\omega_c/\omega_p$,
\end{tabular}
\end{eqnarray*}
where $\omega_c = eB/m_\mathrm{e}$ is the electron cyclotron frequency and $\xi = \lambda_\mathrm{D}/r_{min} = \Gamma^{-3/2}/\sqrt{3}$:
\begin{eqnarray}
\label{eq:Hsp}
\nonumber H(a',\omega',\beta) &=& \frac{8\sqrt{3}}{\pi}\frac{\Gamma^{3/2}}{\omega' a'^2}\int_0^{\xi}\!\!\mathrm{d}k'\int_1^{0}\!\!\mathrm{d}(\cos\theta)\\[2pt]
&\times&\sum_{n=1}^{\infty}nJ_n(k'_\parallel a')^2\left[\mathrm{Im}\frac{-1}{\epsilon(\mathbf{k}',n\omega')}\right].
\end{eqnarray}
Here $J_n(z)$ are Bessel functions and $\epsilon(\mathbf{k}',\omega')$ is the dielectric function re-expressed in terms of scaled, dimensionless arguments,\cite{Matsuda1981}
\begin{eqnarray}
\nonumber \epsilon(\mathbf{k}',\omega') &=& 1 + \frac{1}{{k'}^2}\left[1 + \frac{\omega'}{\sqrt{2}|k'_\parallel|} \sum_{m=-\infty}^{\infty}\exp\left(-\frac{{k'_\perp}^{\!2}}{\beta^2}\right) \right. \\
&\times& \left. I_m\left(\frac{{k'_\perp}^{\!2}}{\beta^2}\right)Z\left(\frac{\omega'+m\beta}{\sqrt{2}|k'_\parallel|}\right) \right].
\end{eqnarray}
The functions $I_m(z)$ and $Z(x_m)$ are the modified Bessel function of the first kind and the plasma dispersion function, respectively.

Eq.~(\ref{eq:Hsp}) reduces to the low-amplitude, high-frequency AC conductivity theory for magnetized plasma presented in Ref.~\onlinecite{ObermanShure1963} and extended for the case of spatially uncorrelated ions by Matsuda.\cite{Matsuda1981} For $a = eE/m_\mathrm{e}\omega_{RF}^2 \ll r_{min}$ we find
\begin{equation}
\label{eq:Hac}
    \lim_{a\ll r_{min}}\!\!\!H = {\omega'}^2\frac{\mathrm{Re}\,\sigma_\parallel(\omega')}{\epsilon_0\omega_p},
\end{equation}
where $\sigma_\parallel$ is the magnetic field-dependent longitudinal AC conductivity given by Matsuda.\cite{Matsuda1981} An analogous procedure could be developed to produce expressions to similarly evaluate heating rates for RF field geometries transverse to or at an arbitrary angle to the magnetic field,\cite{Deutsch2005,Nersisyan2011} however this is outside the scope of this work. The transverse RF field heating rates would be expected to be very different than in the longitudinal case, and so limited information about them can be obtained from the longitudinal heating predictions.

Having derived the expression for the heating rate in Eq.~(\ref{eq:Hsp}), we seek to examine predicted heating rates as a function of $\omega_{RF}$, $E_0$, $\Gamma$ and $B$ to get a general sense of how the heating rate is expected to scale with these parameters as well as generate predicted heating rates that could be measured experimentally by UNPs. We do this by evaluating $H$ as a function of the dimensionless parameters $\beta$, $\omega'$, and $a/r_{min}$ at selected values of $\Gamma$, with our results presented in the following sections. Using the notation $v_{th} = \lambda_\mathrm{D}\omega_p = \sqrt{k_\mathrm{B}T_\mathrm{e}/m_\mathrm{e}}$, we identify $\beta \equiv \omega_c/\omega_p = \lambda_\mathrm{D}/r_c$. The magnetic field strength can be alternatively parameterized by the ratio of length scales $\kappa = r_c/r_{min} = \xi/\beta$. Likewise, the scaled amplitude $a/r_{min}$ can be rewritten $a' = (a/r_{min})/\xi = (v_i/v_{th})/\omega'$.

We evaluate $H$ by numerically integrating (\ref{eq:Hsp}) using a 2D adaptive quadrature algorithm.\cite{Cubature} The relative tolerance for integration convergence for the evaluations presented in this article was set to a part in $10^{4}$. The infinite sums of terms in the integrand and dielectric function are truncated once numerical convergence to within tolerance is established. Additional care must be taken when computing terms depending on the parameter space; e.g. if the argument or order of the $e^{-x}I_m(x)$ term in the dielectric function is very large, then asymptotic expressions can be substituted to make numerical evaluation feasible.\cite{DLMF}

\section{Predicted heating rate vs. magnetic field and oscillation amplitude}

As a first investigation of predicted heating rates relevant for achievable UNP conditions, we survey the magnetic field dependence of the heating for multiple $\omega'$ and $\Gamma$ values. For all of the evaluations presented in this paper we used electron density and temperatures $n_\mathrm{e} = \SI{1.2e13}{\per\meter\cubed}$, $T_\mathrm{e} =$ \SIlist{4.1;12.3;41}{\kelvin}. These respectively correspond to coupling conditions $\Gamma =$ \SIlist{0.15;0.05;0.015}{}; Debye lengths $\lambda_\mathrm{D} =$ \SIlist{40;70;128}{\micro\meter}; and plasma frequency $\omega_p = 2\pi\cdot\SI{31.1}{\mega\hertz}$. The magnetic field strength when $\beta=1$ associated with these conditions is \SI{11.1}{\gauss}.

\begin{figure}
    \centering
    \includegraphics[scale=0.8]{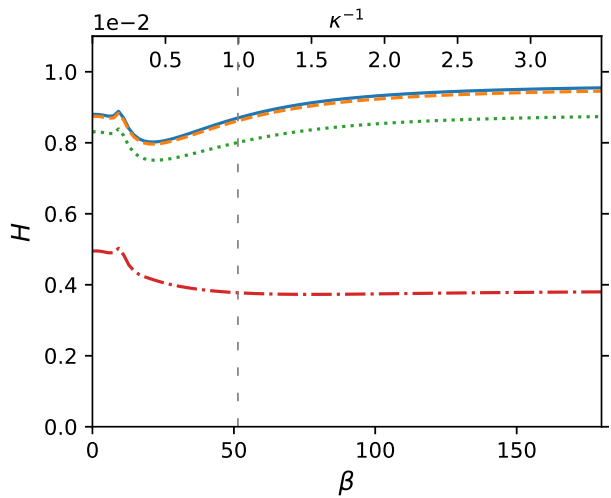}
    \caption{Scaled RF heating rate $H$ as a function of magnetization strength $\beta$ at $\Gamma = 0.05$, RF frequency $\omega' = 10$, and various amplitudes $a/r_{min}$. The solid line is calculated from the low-amplitude limit Eq.~(\ref{eq:Hac}); the dashed, dotted, and dash-dotted lines are evaluations of the finite-amplitude expression Eq.~(\ref{eq:Hsp}) at $a/r_{min} = 1,\,3,\,10$, respectively. The vertical (loosely dashed) line at $\beta = 51.4$ indicates the magnetization strength where $\kappa = r_c/r_{min} = 1$.}
    \label{fig:Fig1}
\end{figure}

The variation of longitudinal AC conductivity heating is predicted to be mild even across large variations in the degree of magnetization of the electrons\cite{Baalrud2017} for many values of $\omega'$ as long as the value of $a/r_{min}$ is not too large.  Fig.~\ref{fig:Fig1}
 shows an example of such a mild variation as a function of $\beta$. This is in stark contrast to e.g. variation in cross-dimensional thermalization rates in magnetized one-component plasmas\cite{Glinsky1992,Beck1996} where such a variation in magnetic field produces orders-of-magnitude variation in thermalization rate. The difference between the two cases that leads to this qualitatively contrasting behavior is that in the longitudinal AC conductivity heating case very high-angle collisions that result in a 180 degree deflection of the electron along a field line contribute to the heating rate.  These collisions do not contribute to a cross-dimensional thermalization rate, however. Thus, substantial heating rates in a longitudinal AC conductivity configuration are still possible even as the electrons become more tightly linked to the magnetic field lines at higher magnetic fields.

\begin{figure*}
    \centering
    \includegraphics[scale=0.8]{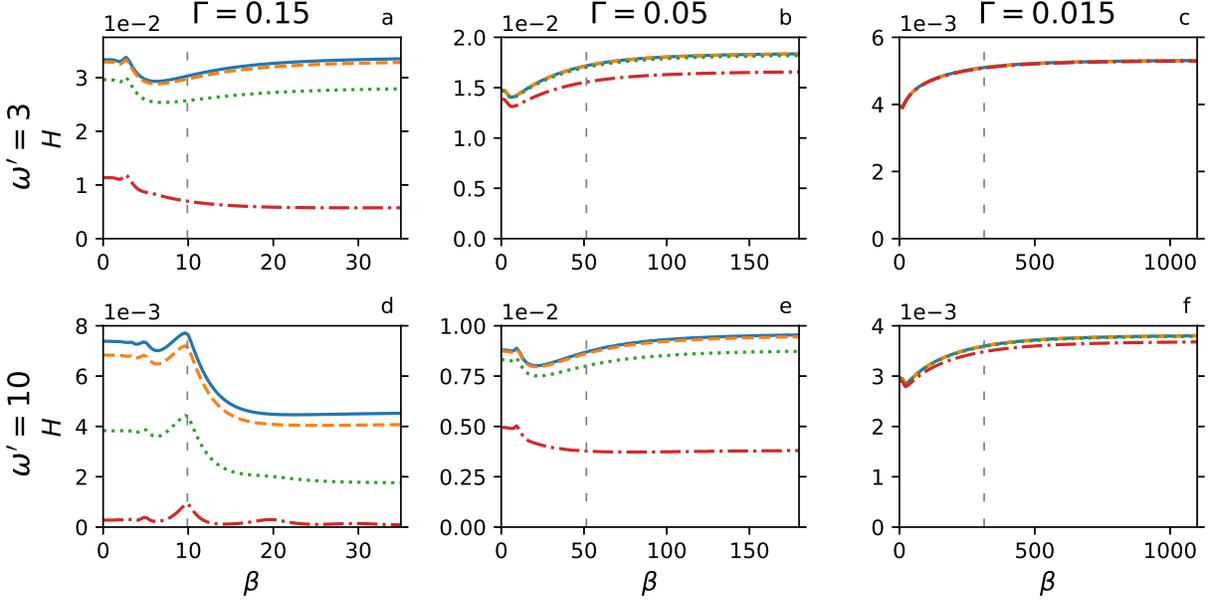}
    \caption{Scaled RF heating rate $H$ as a function of magnetization strength $\beta$ at various amplitudes $a/r_{min}$, same as Fig.~\ref{fig:Fig1}. The top row (a, b, c) corresponds to driving frequency $\omega' = 3$, and the bottom row (d, e, f) has $\omega' = 10$. The three columns correspond to coupling parameters $\Gamma = 0.15, 0.05, 0.015$, respectively. The vertical (loosely dashed) lines once again indicate the $(\Gamma,\beta)$ conditions where $r_c = r_{min}$.}
    \label{fig:Fig2}
\end{figure*}

Fig.~\ref{fig:Fig2}
 shows the same calculations as that in Fig.~\ref{fig:Fig1} for different values of $\omega'$ and $\Gamma$. Mild variation of heating rate with $\beta$ is the most commonly observed feature across most parameters, with the exception being at higher $\omega'$ and higher $\Gamma$. Even in that case the variation is by a factor of two rather than orders-of-magnitude. This relatively small variation in $H$ despite substantial changes in magnetization was commonly observed throughout all the calculations that were performed (with $a/r_{min} < 10$), with only small exceptions near resonances that will be described below. Thus, for the range of $\Gamma=0.015-0.15$, one of the main conclusions of this work is that large variations in magnetization are only expected to produce variations in RF heating rates of order unity at sufficiently small oscillation amplitudes. 

The variations in predicted heating rate with oscillation amplitude are also shown in Figs.~\ref{fig:Fig1} and \ref{fig:Fig2} through the different curves included in those figures. We evaluate (\ref{eq:Hsp}) for values of $a/r_{min} = 1,\,3,\,10$. In addition, a curve providing for a comparison to the predictions from Ref.~\onlinecite{Matsuda1981} in the low-amplitude limit, Eq.~(\ref{eq:Hac}), is shown as well. As expected, there is agreement at the percent to several percent level in the low-amplitude limit between (\ref{eq:Hsp}) and (\ref{eq:Hac}) when $a/r_{min}$ is set to 1 in the Vlasov-Poisson equation-based-predictions.

\begin{figure}
    \centering
    \includegraphics[scale=0.60]{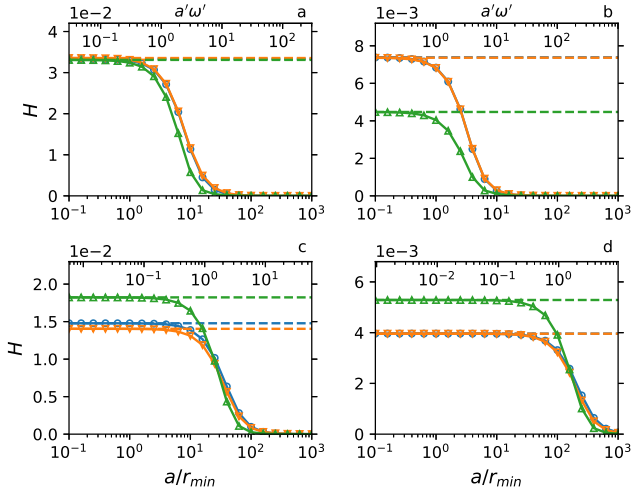}
    \caption{$H$ as a function of amplitude (semi-log) at a) $\Gamma = 0.15,\, \omega'=3$; b) $\Gamma = 0.15,\, \omega'=10$; c) $\Gamma = 0.05,\, \omega'=3$; d) $\Gamma = 0.015,\, \omega'=3$. The lower-horizontal axes are in terms of $a/r_{min}$, and the upper- are $a'\omega'$. The circles are for $\beta = 0$, the downward triangles are at $\Gamma$-dependent $\beta$'s representative of strong magnetization when $r_{min} < r_c < \lambda_\mathrm{D}$, and the upward triangles are at extremely magnetized $\beta$'s when $r_c < r_{min}$ (see text). The dashed lines are the corresponding low-amplitude values.}
    \label{fig:Fig3}
\end{figure}

Fig.~\ref{fig:Fig3}
 shows the dependence of $H$ on the amplitude $a$ in more detail for several conditions. The curves in each panel correspond to evaluations in different $\Gamma$-$\beta$-dependent magnetization regimes\cite{Baalrud2017}: we chose $\beta = 2.5,\,5.7,\,14.1$ for strong magnetization when $r_{min} < r_c < \lambda_\mathrm{D}$ and $\beta = 25,\,130,\,800$ for extreme magnetization when $r_c < r_{min}$ for the three $\Gamma = 0.15,\,0.05,\,0.015$ conditions, respectively (e.g. the curves in Fig.~\ref{fig:Fig3}c are at $\beta = 0,\,5.7,\,130$). The dependence of $H$ on the parameter $a$ is shown with two dimensionless scalings. $H$ decreases substantially as $a$ is increased arbitrarily, which is consistent with the decrease in electron-ion collisions with increasing relative velocity. The onset of $H$ decreasing significantly with $a$ occurs around the condition $a'\omega'\approx1$. This is true across magnetization conditions.\cite{Hu2009} The condition of $a'\omega'=1$ is equivalent to the point where the amplitude of the velocity oscillation is equal to the thermal velocity of the electrons $v_{th}$, and the reduction of $H$ with $a$ at this point is generally expected.\cite{Honda1963,MayCramer1970,Peter1991,Decker1994,NersisyanBook}

While this onset of significant change of $H$ with $a$ occurs at approximately the same condition regardless of the value of $\beta$ for otherwise similar conditions, the structure of the decrease of $H$ with amplitude as $a$ increases past this onset depends on the value of $\beta$. For $\beta$ values corresponding to a strong degree of magnetization or less, we did not observe significant variations in the structure of the decrease of $H$ with $a$ as a function of $\beta$. That was not true between the strong degree of magnetization and the extreme degree of magnetization curves. For those curves, the extreme degree of magnetization curves show a steeper decrease with $a$ than the strongly magnetized curves. The mild variation of $H$ with $\beta$ present at smaller values of $a$ is no longer as uniformly present at larger values of $a$.

\section{Variation of heating rate with applied RF frequency}
Resonance features are visible in Figs.~\ref{fig:Fig1} and \ref{fig:Fig2} for some conditions where $\omega'=\beta$ when the RF frequency is resonant with the cyclotron frequency. While a resonant response is not surprising under these conditions, given the longitudinal geometry such effects would be expected to be much smaller than in a transverse geometry.\cite{Matsuda1981} When the electric field is transverse to the magnetic field, there is a circular polarization that would be stationary in a frame that rotates at the cyclotron frequency leading to a larger and growing system response. In such a transverse geometry the authors of Ref.~\onlinecite{ObermanShure1963} indicate that linear response theory must necessarily breakdown in this case. While the physics causing this breakdown is not as severe in the longitudinal case, the validity of these predictions is less reliable near this resonance condition. However, predictions near resonance conditions are a small fraction of the parameter space that was investigated and these considerations do not alter the general trends calculated away from the resonance condition.

\begin{figure*}
    \centering
    \includegraphics[scale=0.72]{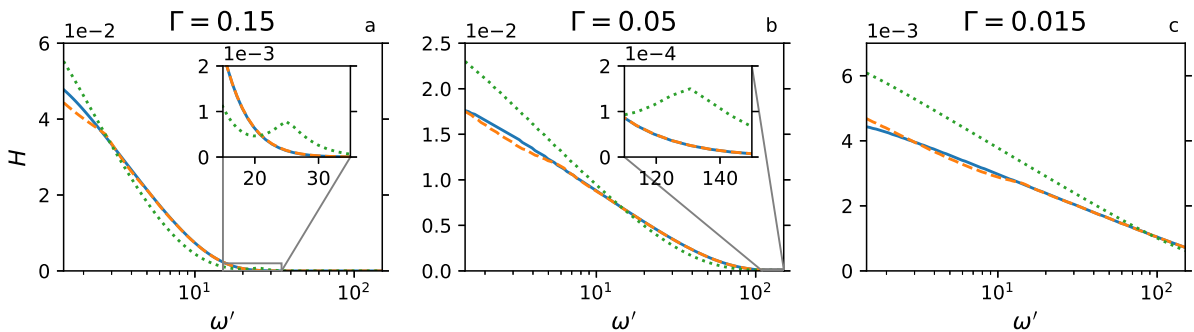}
    \caption{Low-amplitude evaluations of $H$, Eq.~(\ref{eq:Hac}), as a function of $\omega'$ (semi-log) at various magnetization strengths. The three panels (a,b,c) correspond to $\Gamma = 0.15,\, 0.05,\, 0.015$, respectively. The solid lines are the $\beta = 0$ unmagnetized limits; the dashed lines are at $\Gamma$-dependent $\beta$'s representative of strong magnetization; likewise, the dotted lines are at extremely magnetized $\beta$'s (see Fig.~\ref{fig:Fig3} and text). The insets are zoomed in views of the panels around $\omega' = \beta_{r_c < r_{min}}$ and show resonance features similar to those in Figs. \ref{fig:Fig1} and \ref{fig:Fig2}. }
    \label{fig:Fig4}
\end{figure*}

\begin{figure}
    \centering
    \includegraphics[scale=0.62]{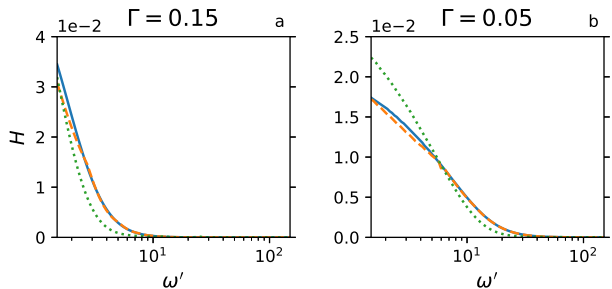}
    \caption{Finite-amplitude evaluations of $H$ as a function of $\omega'$ from Eq.~(\ref{eq:Hsp}) at $a/r_{min} = 10$ and for a) $\Gamma = 0.15$, b) $\Gamma = 0.05$. The solid, dashed, and dotted lines correspond to the various magnetization strengths plotted similarly in Fig.~\ref{fig:Fig4}.}
    \label{fig:Fig5}
\end{figure}

More broadly, in Fig.~\ref{fig:Fig4},
 the variation in $H$ as a function of $\omega'$ is shown for three values of $\Gamma$ and three degrees of magnetization (the solid, dashed, and dotted lines correspond respectively to the same $\beta$ values plotted with circles, downward triangles, and upward triangles in Fig.~\ref{fig:Fig3}). Fig.~\ref{fig:Fig5}  shows the results of similar calculations at a larger oscillation amplitude. The decrease of $H$ with increasing $\omega'$ is consistent with previous predictions.\cite{Matsuda1981,Decker1994} For unmagnetized and strongly-magnetized conditions, only mild variations are observed across different values of $\omega'$ for the values of $\Gamma$ investigated. Increasing the magnetization does produce differences as was noted previously, but across all the conditions studied here with small enough oscillation amplitudes ($a/r_{min} \lessapprox 10$) these differences are relatively minor despite order-of-magnitude changes in degree of magnetization. 

While there are not order-of-magnitude changes as a function of magnetization evident in Figs.~\ref{fig:Fig4} and \ref{fig:Fig5}, there are quantitative differences that suggest fruitful lines of experimental investigation by identifying trends that can be tested. For instance, the rate of change in $H$ with increasing $\omega'$ is predicted to steepen with increasing electron magnetization once the magnetization becomes large enough. Whether or not increasing magnetization is expected to reduce or increase $H$ for a given set of conditions also changes with $\omega'$, $\Gamma$, and $a$ as indicated.

Overall, though, what is most notable across all of the results plotted in Figs.~\ref{fig:Fig4} and \ref{fig:Fig5} is that consistently there is only a mild dependence on magnetic field for appropriately chosen dimensionless parameters. This is despite the fact that the degree of magnetization of the electrons is changing profoundly over the values of $\beta$ that have been investigated. This general observation of only mild dependence on electron magnetization in the predictions of this linear response theory with regard to heating arising from finite AC conductivity is the central one obtained as a result of this work as it informs expectations for measurements associated with UNP laboratory-accessible scales.

\section{Beyond Vlasov-Poisson Equations}

The linear response theory developed from the Vlasov-Poisson equations presented in this work relies on several approximations, and so it is reasonable to inquire how robust the predictions described in the previous sections are. One step towards adding missing physics back in is to include some effect of collisions in equation (1). An initial way of doing so is to include a  Bhatnagar–Gross–Krook (BGK) relaxation approximation term in (\ref{eq:linVP}) as was done in Ref.~\onlinecite{Nersisyan2011}; the collision term on the right-hand side of the equality in (\ref{eq:linVP}) can be approximated by
\begin{equation}
\left(\frac{\partial f_\mathrm{e}}{\partial t}\right)_{coll} = -\gamma\left(f_{\mathrm{e}1} - \frac{n_{\mathrm{e}1}}{n_{\mathrm{e}0}}f_{\mathrm{e}0}\right).
\label{eq:BGK}
\end{equation}
This term captures the fact that collisions will tend to restore the electron distribution $f_\mathrm{e}$ toward equilibrium $f_{\mathrm{e}0}$ at a rate set by the electron collision rate $\gamma$. The collision-inclusive dielectric function $\epsilon(\mathbf{k}',\omega',\gamma)$ can be derived in a number-conserving fashion using this relaxation term.\cite{Nersisyan2011} The collision-inclusive scaled heating rate $H(\beta,\gamma)$ can then be computed by substituting the collisionless $\epsilon$ with the collision-inclusive $\epsilon(\mathbf{k}',\omega',\gamma)$. 

Ref.~\onlinecite{Nersisyan2011} approximates the electron collision rate in the plasma as the sum of electron-electron and electron-ion collision rates $\gamma = \gamma_\mathrm{ee} + \gamma_\mathrm{ei}$ and provides integral expressions to evaluate these rates including dynamical screening effects as a function of magnetic field. Since we are most interested here in making general estimates of effects on $H$, we make a number of simplifications to evaluate $\gamma$. First, we note that the ratio of collision rates $\gamma_\mathrm{ei}/\gamma_\mathrm{ee} \sim (m_\mathrm{e} \ln \Lambda_\mathrm{ei})/(m_\mathrm{i} \ln \Lambda_\mathrm{ee})$ in most unmagnetized UNPs formed via photoionization is much less than one because $m_\mathrm{e}/m_\mathrm{i} \approx 10^5$ and the electron-ion Coulomb logarithm $\ln\Lambda_\mathrm{ei}$ is typically within one to two orders of magnitude of $\ln\Lambda_\mathrm{ee}$. We also use the low-temperature static screening approximation to estimate the electron-electron collision rate at $B = 0$ as\cite{Spitzer1962,Ichimaru1970,Nersisyan2011}
\begin{equation}
\gamma_\mathrm{ee} \approx \dfrac{4\sqrt{2\pi}}{3}\dfrac{e^4n_\mathrm{e}}{(4\pi\epsilon_0)^2m_\mathrm{e}^2v_{th}^3}\ln\xi.
\end{equation}
Furthermore, evaluations of $\ln\Lambda_\mathrm{ee}$ as a function of magnetic field predict a bounded behavior.\cite{Nersisyan2011} For conditions corresponding to $\xi=10$, with increasing magnetic field the Coulomb logarithm comes to a maximum of within twice the field-free value before diminishing to half of the field-free value as the magnetic field approaches infinite strength. Therefore we can approximate the electron collision rate $\gamma$ at arbitrary magnetic field strength as a range of factors of two around the $\Gamma$-dependent, magnetic-field free value $\gamma_0 \approx \gamma_\mathrm{ee}$. Given the level of approximation involved in the replacement of the collision term with the BGK term in (\ref{eq:BGK}), a more precise determination of $\gamma$ is not very meaningful.

\begin{figure}
    \centering
    \includegraphics[scale=0.73]{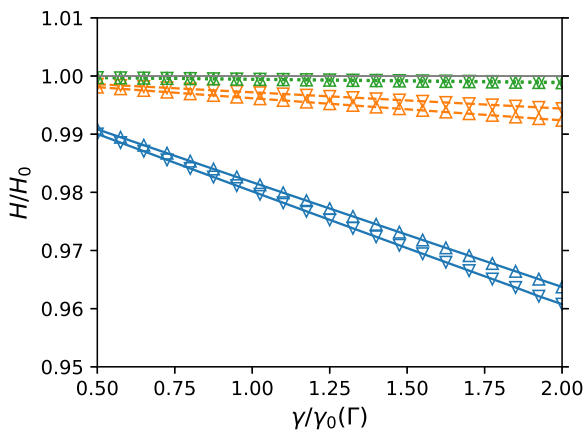}
    \caption{Ratio of scaled heating rates computed at $a' = 1/3,\,\omega' = 3$ with a collision-inclusive ($H$) and collisionless ($H_0$) dielectric response as a function of electron collision rate $\gamma$ for various couplings and magnetizations. The lines in solid (blue), dashed (orange), and dotted (green) are for $\Gamma =$ \SIlist{0.15;0.05;0.015}{}, respectively; the lines with downward and upward pointing triangles are at the respective strong and extreme magnetization strengths (see Fig.~\ref{fig:Fig3} caption and text). The collision rate is shown scaled to the coupling-dependent, magnetic-field-free rate $\gamma_0(\Gamma) = \gamma_\mathrm{ee}$. Note that the y-axis does not begin at 0.}
    \label{fig:Fig6}
\end{figure}

Using this range for the collisional relaxation rate, the impact on the predicted value of $H$ was of order or less than 5\% for a range of magnetization and coupling conditions evaluated at $a'\omega' = 1$, as seen in Fig.~\ref{fig:Fig6}. This is yet another way that in the context of the calculations presented the dependence of predictions on $\beta$ was found to be mild for sufficiently small driving amplitudes.

This relaxation approximation will be most effective for lower values of $\Gamma$ and lower values of $\beta$. Including a more complete description of collisions in the determination of $H$ occurring in the context of a high degree of magnetization\cite{Baalrud2020} is beyond the scope of this work. It is reasonable to expect, though, that for the parameters investigated here where $\Gamma$ remains much less than one, the mild sensitivity to magnetization that is the main feature is likely to still be the case. At the same time, quantitative differences could well be observed in a more sophisticated treatment for higher $\Gamma$ values and higher $\beta$ values where the overall formalism is more suspect.

It is also reasonable to consider the trustworthiness of the higher oscillation amplitude predictions. At some point, the linearization of the Vlasov equation will no longer be valid. A related question is how any such breakdown would vary as function of magnetic field. We have deliberately limited the x-axes extents in Figs.~\ref{fig:Fig3} and \ref{fig:Fig5} to correspond to regions where the linearization is still reasonable, where reasonable is defined as the amount of heating for the electrons from the RF in a collision time being less than the average thermal energy. For the unmagnetized case, the numerical predictions in Ref.~\onlinecite{Decker1994} are consistent with this being a reasonable choice of limit. The lack of variation in $H$ with magnetization indicates similar ranges of validity across values of $\beta$. Anisotropic temperature distributions may develop, but extensions to the theory presented here to include anisotropic temperature distributions shows that such effects are mild, being typically several percent.

There are experimental measurements which are still under final analysis that will be presented by the authors in future work that are conducted under conditions similar to those presented in this work. These measurements seem broadly consistent with this general observation in that order-of-magnitude changes in magnetization do not produce order-of-magnitude changes in $H$ and that measured values of $H$ at a high degree of magnetization can be close to those at lower values of magnetization, while some quantitative differences are very likely present.

\section{Conclusion}

We have used linear response theory based on the Vlasov and Poisson equations to predict heating rates arising from the finite AC conductivity in ultracold neutral plasmas when these plasmas are subjected to oscillating electric fields. This work extends similar treatments\cite{Matsuda1981,Decker1994} to include both the effects from magnetization of the electrons and effects from finite oscillation amplitudes. Predictions were developed in a longitudinal geometry where the oscillating electric field is aligned along the direction of the applied magnetic field. For sufficiently small oscillation amplitudes, agreement with the low-amplitude, uncorrelated ion theory of Ref.~\onlinecite{Matsuda1981} was obtained. The deviation from this agreement with increasing oscillation amplitude was characterized and quantified. 

A survey across common low-density UNP conditions showed that for detectable but small oscillation amplitudes, only mild variation of heating rate coefficients is predicted despite order-of-magnitude changes in the degree of electron magnetization. This is in sharp contrast to measurements of cross-dimensional thermalization, for instance.\cite{Glinsky1992,Beck1996} For many conditions the variation is less than 10s of percent across the full range of applied magnetic field. For others, factor of 2 or so variations are predicted to occur. This theory can be used to determine parameters where these larger degrees of variation are predicted to exist, guiding investigations with more sophisticated theories or experimental measurements. The theory presented here serves not only to provide predictions for heating rates under magnetized UNP conditions, but also indicates useful parameters for conducting measurements in a wide parameter space as well. In addition to some measurements currently in the final phase of analysis, such measurements are planned to be conducted in the future.

\begin{acknowledgments}
This work was supported by the Air Force Office of Scientific Research (AFOSR), Grant No. FA9550-17-1-0148.
\end{acknowledgments}

\section*{Data availability}
The data that support the findings of this study are available from the corresponding author upon reasonable request.

\bibliography{JGuthrie_JRoberts}

\end{document}